 \documentclass[twocolumn,showpacs,showkeys,preprintnumbers,amsmath,amssymb]{revtex4} 
\usepackage{graphicx}% Include figure files
\usepackage{dcolumn}% Align table columns on decimal point
\usepackage{bm}% bold math
\begin{document}

\title{Robust quantum dot state preparation via adiabatic passage with frequency-swept optical pulses}

\author{C.-M. Simon$^{1,2}$}
\author{T. Belhadj$^1$}
\author{B. Chatel$^2$}
\author{T. Amand$^1$}
\author{P. Renucci$^1$}
\author{A. Lemaitre$^3$}
\author{O. Krebs$^3$}
\author{P. A. Dalgarno$^{4}$}
\author{R. J. Warburton$^{4,5}$}
\author{X. Marie$^1$}
\author{B. Urbaszek $^{1}$}
\email[Corresponding author : ]{urbaszek@insa-toulouse.fr}

\affiliation{%
$^1$Universit\'e de Toulouse, INSA-CNRS-UPS, LPCNO, 135 Av. Rangueil, 31077 Toulouse, France}

\affiliation{%
$^2$Universit\'e de Toulouse, CNRS, LCAR, IRSAMC, 118 Rt. Narbonne, 31062 Toulouse, France}

\affiliation{%
$^3$Laboratoire de Photonique et Nanostructures CNRS, route de Nozay, 91460 Marcoussis, France}

\affiliation{%
$^4$ School of Engineering and Physical Sciences, Heriot-Watt University, Edinburgh, EH14 4AS, UK}

\affiliation{%
$^5$ Department of Physics, University of Basel, Klingelbergstrasse 82, 4056 Basel, Switzerland}

\date{\today}

\begin{abstract}
The energy states in semiconductor quantum dots are discrete as in atoms, and quantum states can  be coherently controlled with resonant laser pulses. Long coherence times allow the observation of Rabi-flopping of a single dipole transition in a solid state device, for which occupancy of the upper state depends sensitively on the dipole moment and the excitation laser power. We report on the robust preparation of a quantum state using an optical technique that exploits rapid adiabatic passage from the ground to an excited state through excitation with laser pulses whose frequency is swept through the resonance. This observation in photoluminescence experiments is made possible by introducing a novel optical detection scheme for the resonant electron hole pair (exciton) generation. 

\end{abstract}

\pacs{73.21.La,78.55.Cr,78.67.Hc}% PACS, the Physics and Astronomy
                             % Classification Scheme
                             % 78.55.Cr   III-V semiconductors  
                             % 73.21.La    Quantum dots   
                             % 78.67.Hc  Optical properties 
                            \keywords{Quantum dots}%Use showkeys class option if keyword
                             %display desired
\maketitle

Photon correlation measurements along with resonant laser scattering have established the atom-like character of the interband transitions in quantum dots \cite{Michler2000,Atature2009,Flagg2009}.
Excitation of a two level system by a short, intense laser pulse can induce an oscillation of the system between the upper and lower state during the pulse, at the Rabi frequency $\Omega(t)=\mu A(t)/\hbar$ where $\mu$ is the dipole moment of the transition and $A(t)$ the electric field envelope of the laser pulse. Any quantum state (qubit) manipulation scheme benefits from the long coherence times in quantum dots \cite{Borri2001} and necessitates fast and robust initial state preparation \cite{Steel2001,Steel2003,press08}. In principle a two level system can be initialised in the upper state with a maximum fidelity of 100\% if the laser power is optimised in order to induce exactly half a Rabi oscillation during the pulse duration, a so-called $\pi$ pulse \cite{Kamada2001,Zrenner2002,Ramsay2010}. Although Rabi oscillations observed in a single dot or for individual atoms \cite{Browaeys2009} are a beautiful example of strong coupling between laser light and a single dipole, this commonly used technique presents two major drawbacks: (i) the upper state population is highly sensitive to fluctuations in the system, such as laser power, and (ii) in measurements on dot ensembles, an inhomogeneous distribution of dipole moments and transition energies among dots requires different laser intensities and frequencies for inducing a population transfer of the dot ensemble.

\begin{figure}
\includegraphics[width=0.46\textwidth]{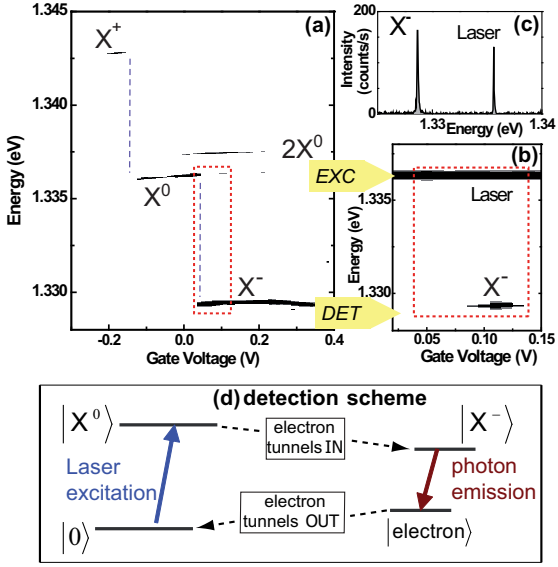}
\caption{\label{fig:fig1} (a)  Contour plot of the PL intensity at 4K as a function of gate voltage applied to the sample for non-resonant excitation in the wetting layer. white $<$ 50 counts, black $\ge$ 1000 counts. The positively charged exciton X$^+$, and the neutral exciton X$^0$, biexciton 2X$^0$ and the negatively charged exciton X$^-$ were identified through gate voltage dependence and fine structure analysis (b) Contour plot PL as function of voltage with a narrowband cw laser resonant with the  X$^0$ transition. The X$^-$ transition appears for a gate voltage range of about 50mV. (c) spectrum when the laser is resonant with the  X$^0$ transition leading to X$^-$ emission. (d) Energy level diagram with neutral exciton X$^0$ absorption monitored via charged exciton X$^-$ emission.  
}
\end{figure}

Here we show that these drawbacks can be overcome by inducing an adiabatic passage with frequency-swept laser pulses. Unlike Rabi cycling, adiabatic passage is robust against small-to-moderate variations in the laser intensity, detuning, dipole moment and interaction time \cite{Vitanov2001}. Chirped radiofrequency (RF) pulses are commonly used in nuclear magnetic resonance (NMR) based imaging \cite{Garwood2001}.
In the context of semiconductor quantum dots adiabatic population transfer has been induced by a slow variation of the electrostatically defined confinement potential \cite{Petta2010}.
 For applications in atomic physics and chemistry for quantum states separated by frequencies in the optical domain, chirped laser pulses, in strong analogy to NMR, have been used to induce complete population transfer via adiabatic passage 
 \cite{Melinger1994,Vitanov2001,Maas99Rbadiabatic}.
 Our experiments demonstrate the power of adiabatic population transfer between individual quantum states in condensed matter, paving the way for stimulated Raman adiabatic passage (STIRAP)\cite{Vitanov2001,Fleischhauer2005}, efficient spin state preparation \cite{Greilich2006,Eble2009} and quantum condensation of excitons in a microcavity \cite{Eastham2009}.

The two level system investigated here is comprised of a ground state and neutral exciton X$^0$ state of an individual InAs quantum dot in a GaAs matrix embedded in a Schottky diode structure with a 25nm tunnel barrier as in \cite{urbaszek2003}. The experiments are carried out at 4K in a home built confocal microscope built around attocube nano-positioners connected to a spectrometer and a Si charge coupled device camera. Application of a gate voltage enables deterministic loading of individual electrons. To minimise the laser stray light on the detector we employ dark field techniques and linear cross polarization. 

Pioneering work on Rabi oscillations in single semiconductor quantum dots used differential transmission \cite{Steel2001, Steel2003} and photocurrent techniques \cite{Zrenner2002,Ramsay2010} for detection following resonant optical excitation. 
To probe resonant exciton generation in emission poses the problem of laser stray light at the detection wavelength reaching the detector. Implementation of up-conversion techniques \cite{Marie01}, wave guide sample structures \cite{Flagg2009} or non-resonant emitter - cavity coupling \cite{Michler2009} have allowed this problem to be circumvented.
Using charge tuneable structures allowed us to implement a novel detection scheme that does not suffer from a strong background signal as photocurrent. The generated X$^0$ population is probed by monitoring the negatively charged exciton X$^-$ emission, which plays the role of a spectator state at lower energy, see energy level diagram in figure 1d. In a small window of voltage, at a time $\tau_\text{in}$ (which we estimate to be  $\approx 70ps$ \cite{tunnel})  the photogenerated X$^0$ becomes an X$^-$ by grabbing an electron from the Fermi sea in the sample back contact. Radiative emission of the X$^-$ leaves behind a single electron which tunnels out of the dot, allowing again X$^0$ absorption \cite{Dalgarno2008b}. 

\begin{figure}
\includegraphics[width=0.45\textwidth]{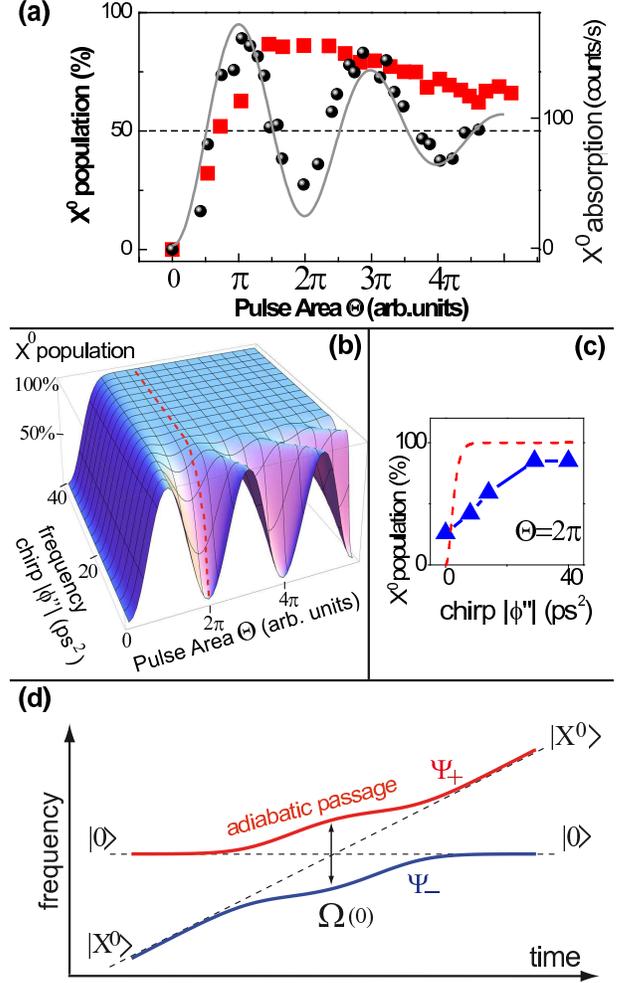}
\caption{\label{fig:fig2} (a) Rabi oscillations (black circles data point, gray line is a fit with model adapted from \cite{Ramsay2010}) for excitation with Fourier transform (FT) limited pulse and adiabatic passage for strongly chirped laser with $\phi'' = -40~ps^2$ (red squares). (b) Numerical calculation (see Eq. \ref{eq:Liouville}) of the X$^0$ population as a function of laser pulse area $\Theta$ for different chirps up to $|\phi''| = 40~ps^2$, the maximum value used in the experiment. The presented graph is symmetrical with respect to a change in sign of  $\phi''$ . Dashed red line shows X$^0$ population for constant $\Theta=2\pi$. (c) Blue triangles: experimentally achieved X$^0$ population as a function of frequency chirp for a $\Theta=2\pi$ pulse, dashed red line (same as in (b)) shows calculated X$^0$ population for constant $\Theta=2\pi$. (d) Eigenfrequencies of the dressed states $\lvert \Psi_+(t) \rangle$ and $\lvert \Psi_-(t) \rangle$ given by $1/2[-\Delta(t)\pm\sqrt{\Delta(t)^2+\Omega(t)^2}]$ (see Hamiltonian in Eq. \ref{eq:hamiltonian}) calculated as a function of time for $\Theta=5\pi$ and $\phi'' = - 40 ~ ps^2 $.  
}
\end{figure}

In a first step, the population transfer achieved via Rabi cycling between the ground state $\lvert 0 \rangle$ and the exciton state $\lvert X^0 \rangle$ will serve as a reference for the robust exciton generation via an adiabatic passage for the same quantum dot.
The lifetime $T_1$ and the maximum coherence time $T_2=2T_1$ of the optically created X$^0$ are in our case limited by the tunnelling time $\tau_\text{in}$. To observe Rabi oscillations, $\tau_\text{in}$ should be longer than the pulse duration $\tau_0 \simeq 3 ps$ (FWHM) of spectral width $\simeq0.5 meV$ from the pulsed Ti-Sa laser. 
The photogenerated X$^0$ population undergoes clear sinusoidal oscillations when plotted versus the pulse area $\Theta =\int_{-\infty}^{t}\Omega(t')d t'$  where $t\gg \tau_0$ \cite{Steel2001}, see figure 2a. This time integrated Rabi frequency is proportional to the square root of the average laser intensity. The strictly resonant excitation in our experiment leads without damping to oscillations of the X$^0$ population between 0 and 100\% \cite{press08}. A first maximum for the X$^0$ population of about 90\% is observed for a pulse area equivalent to the rotation angle of the system of $\Theta=\pi$ (a $\pi$  pulse) see figure 2a. The contrast of the Rabi oscillations does not change significantly over the bias voltage range (0.07 to 0.12V) which results in X$^-$ emission upon X$^0$ excitation. As $\tau_\text{in}$ does change when the tunnel barrier height is changed, this observation provides another indication that $\tau_\text{in} \gg \tau_0 $. Symmetric damping of the Rabi oscillations as a function of pulse area indicates an excitation power dependent dephasing process, as for example the interaction with acoustic phonons \cite{Ramsay2010}. The gray line in figure 2a is based on the model in \cite{Ramsay2010} i.e. allowing for transitions between the two dressed states (see below), taking into account that in our experiments the thermal energy is comparable to or smaller than the Rabi splitting $\hbar \Omega$. The comparison between these calculations and the data serve to express the X$^0$ population in percent.

Using the same quantum dot and detection techniques presented in figure 1, we present in what follows a successful implementation of adiabatic passage to prepare robustly a single quantum dot in an exciton state X$^0$. 
The key change lies in the excitation pulse: the system is excited \emph{resonantly} by a laser pulse that passed a compressor comprised of two reflection gratings with 2000 lines/mm \cite{chatelNa03PRA} before reaching the sample. As a result, the frequency in the excitation pulse sweeps now linearly in time \cite{Melinger1994}, with an instantaneous pulse frequency $\omega_i(t)=\omega_0+2bt$, where $b$ is the sweep rate.  
The frequency sweep should be slow enough such that the system is at any given time during the interaction in an eigenstate \cite{Melinger1994}. 
Results for the X$^0$ absorption as a function of pulse area are plotted on figure \ref{fig:fig2}a. 
Excitation with a Fourier transform limited pulse results in Rabi oscillations, whereas excitation with a chirped pulse is strikingly different, showing no oscillations of the prepared X$^0$ population.
For the pulse area corresponding to a $\pi$ Rabi pulse, the chirped pulse does not yet create an efficient population transfer $\lvert 0 \rangle \to \lvert X^0 \rangle$. This is achieved for a pulse area $\ge 1.5\pi$, beyond which the created X$^0$ population changes very little with increasing power \cite{pulsearea}. This is a clear signature of a robust population inversion via an adiabatic process. 
Adiabatic passage is a coherent process. The disappearance of the Rabi oscillations due to \emph{decoherence} would result in a pulse area (laser power) dependence in figure 2a that would slowly approach from below, but never exceed a 50\% X$^0$ population probability, as verified experimentally in measurements using a cw diode Laser for excitation.
The adiabatic passage data in figure 1a shows for $\Theta\ge 1.5\pi$ for all data points an X$^0$ probability above 60\%. The exact reason why the generated X$^0$ population seems to decrease from the initial 90\% when using higher laser power is at present unknown, it would be interesting for the future to develop an acoustic phonon interaction based model similar to the work of Ramsay et al\cite{Ramsay2010}, but for adiabatic processes \cite{asy}.

To further demonstrate the influence of the laser chirp on the X$^0$ population, we plot in figure 2c the measured X$^0$ population created for a $2\pi$ pulse, which for an undamped two level system is zero for zero chirp, as a function of the applied chirp. The lowest value of only about 25\% for the damped Rabi oscillations rises gradually to about 90\% for an adiabatic passage with maximum chirp, with intermediate values for smaller chirps.

Figure 2b shows a calculation of the amount of chirp necessary to pass from the Rabi oscillation regime to robust adiabatic passage in the absence of damping. The electric dipole interaction of the laser pulse with the two level system $\lvert 0 \rangle$ and $\lvert X^0 \rangle$ separated in energy by $\hbar \omega_0$ is:
\small
\begin{equation}
\label{eq:dipole}
W(t)=-D E(t) = -\mu E(t)\left(\lvert 0 \rangle \langle X^0 \rvert + \lvert X^0 \rangle \langle 0 \rvert\right)
\end{equation}
\normalsize
The real electric field of the excitation pulse $E(t)$ can be written as:
\small
\begin{equation}
\label{eq:efield}
E(t) = \frac{1}{2}\left(A(t) e^{-i\omega_0 t}+A^*(t) e^{i\omega_0 t}\right)
\end{equation}
\normalsize
where $\omega_0$ is the laser frequency in the case of our resonant experiment. In what follows we assume a Gaussian pulse shape \cite{hyper}:
\small
\begin{equation}
\label{eq:At}
A(t)=\frac{\Omega(t)}{2\mu}=\frac{\Theta}{2\sqrt{\pi}\mu \sqrt{\Gamma}}\text{exp}\left(-\Gamma t^2\right)
\end{equation}
\normalsize

$\tau_0$ corresponds to the Fourier limited FWHM pulse duration in intensity while $\phi''$ is the second derivative of the spectral phase of the electric field at the laser frequency and $\Gamma=\frac{\tau_0^2}{2\ln2}-2i\phi''$. Due to the chirp $\phi''$ the excitation frequency sweeps linearly in time during the pulse with a rate $b=\frac{8 \phi'' (ln2)^2}{\tau_0^4+16(\phi'')^2 (ln2)^2 }$. 
This leads for the system-laser interaction to a time-dependent detuning $\Delta(t)=2bt$ assuming the laser is resonant at $t=0$. 
In our setup the double grating compressor introduces a negative chirp $\phi''$ proportional to the distance $d$ between the two gratings in the compressor \cite{Melinger1994}. The adiabatic passage is carried out with $d$ large enough to produce a chirp as strong as $\phi'' = - 40~ps^2 $ leading to a stretched pulse of $\tau \simeq 40$ps, where the chirped pulse width is given by $\tau^2 = \tau_0^2 + \left(\frac{4 \phi'' \text{ln}2}{\tau_0}\right) ^2$.

The Hamiltonian in the frequency modulation frame using the rotating wave approximation, applicable for the laser powers used here, may be written as \cite{Melinger1994}:   
\small
\begin{equation}
\label{eq:hamiltonian}
H=\hbar
\left( \begin{array}{cc}
0 & \frac{\Omega (t)}{2} \\
\frac{\Omega^* (t)}{2} & -\Delta(t) \\
\end{array} \right)
\end{equation}
\normalsize

To evaluate the X$^0$ population as a function of time, that corresponds to the element $\rho_{11}(t)$ of the $2\times2$ density matrix $\rho (t)$,  we solve numerically the Liouville equation: 
\small
\begin{equation}
\label{eq:Liouville}
i\hbar \frac{d\rho(t)}{dt}=[H,\rho (t)]
\end{equation}
\normalsize

In figure 2b we plot the solutions of equation \ref{eq:Liouville} for a time $t > \tau$ varying the pulse area $\Theta$ and the frequency chirp $\phi''$. For the picosecond pulse used in our experiment a chirp $|\phi''|>20ps^2$ of positive or negative sign, achieves robust population transfer via adiabatic passage in our non-damped calculation. 
For example, the calculated X$^0$ population as a function of $\phi''$ for $\Theta=2\pi$ shows the same tendency as the experiments for different values of $|\phi''|$ in figure 2c.
In theory there is no upper limit for a useful $|\phi''|$, in practice an excessively strong chirp (long pulse duration) should be avoided to ensure the system remains coherent during the light-matter interaction \cite{Borri2001}.

An intuitive interpretation of the observed adiabatic passage comes from the dressed state picture \cite{Cohen,Atature2009,Flagg2009}. 
The eigenstates, $\lvert \Psi_+(t) \rangle$ and $\lvert \Psi_-(t) \rangle$, and the corresponding eigenenergies of the two level system coupled to an electromagnetic wave are obtained by diagonalizing equation \ref{eq:hamiltonian}.
During an adiabatic passage the system evolves slowly enough so it does not change eigenstate population. It is the \textit{composition} of the eigenstate that changes as the laser frequency sweeps past resonance. In our case the sweep rate $b<0$, so an adiabatic passage is possible following the dressed state $\lvert \Psi_+(t) \rangle$ in figure \ref{fig:fig2}d. Asymptotically, for times long before and after the pulse, each dressed state becomes uniquely identified with a single unperturbed state
and complete adiabatic inversion $\lvert 0 \rangle \to \lvert X^0 \rangle$ occurs. Considering the time evolution, critical points occur when the two dressed states are close in energy so diabatic transitions are possible \cite{Hioe1984}.

In future experiments it would be intriguing to verify if controlled, resonant exciton generation improves the fidelity of single photon interference experiments \cite{Bennett2009}. Another interesting extension of the presented measurements is biexciton creation via an adiabatic passage to allow controlled generation of entangled photon pairs \cite{Akopian2006,Dousse2010}. The robustness of the adiabatic passage presented here is a starting point for optical manipulation of qubits based on quantum dots coupled through tuneable tunnel barriers \cite{Zoller2003,Hohenester2003} or for a single dot in a transverse magnetic field \cite{Steel2004}. In both scenarios a $\Lambda$ system is formed for which STIRAP can be envisaged \cite{Vitanov2001,Fleischhauer2005}. 

This work was supported by ANR P3N QUAMOS, DGA, PPF LUGSO, IUF, ITN Spinoptronics.
We thank Andrew Ramsay, Cathie Ventalon and Paul Voisin for fruitful discussion.

%% Put the bibliography here, most people will use BiBTeX in
%% which case the environment below should be replaced with
%% the \bibliography{} command.

\begin{thebibliography}{37}
\expandafter\ifx\csname natexlab\endcsname\relax\def\natexlab#1{#1}\fi
\expandafter\ifx\csname bibnamefont\endcsname\relax
  \def\bibnamefont#1{#1}\fi
\expandafter\ifx\csname bibfnamefont\endcsname\relax
  \def\bibfnamefont#1{#1}\fi
\expandafter\ifx\csname citenamefont\endcsname\relax
  \def\citenamefont#1{#1}\fi
\expandafter\ifx\csname url\endcsname\relax
  \def\url#1{\texttt{#1}}\fi
\expandafter\ifx\csname urlprefix\endcsname\relax\def\urlprefix{URL }\fi
\providecommand{\bibinfo}[2]{#2}
\providecommand{\eprint}[2][]{\url{#2}}

\bibitem[{\citenamefont{Michler et~al.}(2000)\citenamefont{Michler, Kiraz,
  Becher, Schoenfeld, Petroff, Zhang, Hu, and Imamoglu}}]{Michler2000}
\bibinfo{author}{\bibfnamefont{P.}~\bibnamefont{Michler}},
  \bibinfo{author}{\bibfnamefont{A.}~\bibnamefont{Kiraz}},
  \bibinfo{author}{\bibfnamefont{C.}~\bibnamefont{Becher}},
  \bibinfo{author}{\bibfnamefont{W.~V.} \bibnamefont{Schoenfeld}},
  \bibinfo{author}{\bibfnamefont{P.~M.} \bibnamefont{Petroff}},
  \bibinfo{author}{\bibfnamefont{L.}~\bibnamefont{Zhang}},
  \bibinfo{author}{\bibfnamefont{E.}~\bibnamefont{Hu}}, \bibnamefont{and}
  \bibinfo{author}{\bibfnamefont{A.}~\bibnamefont{Imamoglu}},
  \bibinfo{journal}{Science} \textbf{\bibinfo{volume}{290}},
  \bibinfo{pages}{2282} (\bibinfo{year}{2000}).

\bibitem[{\citenamefont{Vamivakas et~al.}(2009)\citenamefont{Vamivakas, Zhao,
  Lu, and Atat\"ure}}]{Atature2009}
\bibinfo{author}{\bibfnamefont{N.~A.} \bibnamefont{Vamivakas}},
  \bibinfo{author}{\bibfnamefont{Y.}~\bibnamefont{Zhao}},
  \bibinfo{author}{\bibfnamefont{C.-Y.} \bibnamefont{Lu}}, \bibnamefont{and}
  \bibinfo{author}{\bibfnamefont{M.}~\bibnamefont{Atat\"ure}},
  \bibinfo{journal}{Nature Phys.} \textbf{\bibinfo{volume}{5}},
  \bibinfo{pages}{198} (\bibinfo{year}{2009}).

\bibitem[{\citenamefont{Flagg et~al.}(2009)\citenamefont{Flagg, Muller,
  Robertson, Founta, Deppe, Xiao, Ma, Salamo, and Shih}}]{Flagg2009}
\bibinfo{author}{\bibfnamefont{E.~B.} \bibnamefont{Flagg}},
  \bibinfo{author}{\bibfnamefont{A.}~\bibnamefont{Muller}},
  \bibinfo{author}{\bibfnamefont{J.~W.} \bibnamefont{Robertson}},
  \bibinfo{author}{\bibfnamefont{S.}~\bibnamefont{Founta}},
  \bibinfo{author}{\bibfnamefont{D.~G.} \bibnamefont{Deppe}},
  \bibinfo{author}{\bibfnamefont{M.}~\bibnamefont{Xiao}},
  \bibinfo{author}{\bibfnamefont{W.}~\bibnamefont{Ma}},
  \bibinfo{author}{\bibfnamefont{G.~J.} \bibnamefont{Salamo}},
  \bibnamefont{and} \bibinfo{author}{\bibfnamefont{C.~K.} \bibnamefont{Shih}},
  \bibinfo{journal}{Nature Phys.} \textbf{\bibinfo{volume}{5}},
  \bibinfo{pages}{203} (\bibinfo{year}{2009}).

\bibitem[{\citenamefont{Borri et~al.}(2001)\citenamefont{Borri, Langbein,
  Schneider, Woggon, Sellin, Ouyang, and Bimberg}}]{Borri2001}
\bibinfo{author}{\bibfnamefont{P.}~\bibnamefont{Borri}},
  \bibinfo{author}{\bibfnamefont{W.}~\bibnamefont{Langbein}},
  \bibinfo{author}{\bibfnamefont{S.}~\bibnamefont{Schneider}},
  \bibinfo{author}{\bibfnamefont{U.}~\bibnamefont{Woggon}},
  \bibinfo{author}{\bibfnamefont{R.~L.} \bibnamefont{Sellin}},
  \bibinfo{author}{\bibfnamefont{D.}~\bibnamefont{Ouyang}}, \bibnamefont{and}
  \bibinfo{author}{\bibfnamefont{D.}~\bibnamefont{Bimberg}},
  \bibinfo{journal}{Phys. Rev. Lett.} \textbf{\bibinfo{volume}{87}},
  \bibinfo{pages}{157401} (\bibinfo{year}{2001}).

\bibitem[{\citenamefont{Stievater et~al.}(2001)\citenamefont{Stievater, Li,
  Steel, Gammon, Katzer, Park, Piermarocchi, and Sham}}]{Steel2001}
\bibinfo{author}{\bibfnamefont{T.~H.} \bibnamefont{Stievater}},
  \bibinfo{author}{\bibfnamefont{X.}~\bibnamefont{Li}},
  \bibinfo{author}{\bibfnamefont{D.~G.} \bibnamefont{Steel}},
  \bibinfo{author}{\bibfnamefont{D.}~\bibnamefont{Gammon}},
  \bibinfo{author}{\bibfnamefont{D.~S.} \bibnamefont{Katzer}},
  \bibinfo{author}{\bibfnamefont{D.}~\bibnamefont{Park}},
  \bibinfo{author}{\bibfnamefont{C.}~\bibnamefont{Piermarocchi}},
  \bibnamefont{and} \bibinfo{author}{\bibfnamefont{L.~J.} \bibnamefont{Sham}},
  \bibinfo{journal}{Phys. Rev. Lett.} \textbf{\bibinfo{volume}{87}},
  \bibinfo{pages}{133603} (\bibinfo{year}{2001}).

\bibitem[{\citenamefont{Li et~al.}(2003)\citenamefont{Li, Wu, Steel, Gammon,
  Stievater, Katzer, Park, Piermarocchi, and Sham}}]{Steel2003}
\bibinfo{author}{\bibfnamefont{X.}~\bibnamefont{Li}},
  \bibinfo{author}{\bibfnamefont{Y.}~\bibnamefont{Wu}},
  \bibinfo{author}{\bibfnamefont{D.}~\bibnamefont{Steel}},
  \bibinfo{author}{\bibfnamefont{D.}~\bibnamefont{Gammon}},
  \bibinfo{author}{\bibfnamefont{T.~H.} \bibnamefont{Stievater}},
  \bibinfo{author}{\bibfnamefont{D.~S.} \bibnamefont{Katzer}},
  \bibinfo{author}{\bibfnamefont{D.}~\bibnamefont{Park}},
  \bibinfo{author}{\bibfnamefont{C.}~\bibnamefont{Piermarocchi}},
  \bibnamefont{and} \bibinfo{author}{\bibfnamefont{L.~J.} \bibnamefont{Sham}},
  \bibinfo{journal}{Science} \textbf{\bibinfo{volume}{301}},
  \bibinfo{pages}{809} (\bibinfo{year}{2003}).

\bibitem[{\citenamefont{Press et~al.}(2008)\citenamefont{Press, Ladd, Zhang,
  and Yamamoto}}]{press08}
\bibinfo{author}{\bibfnamefont{D.}~\bibnamefont{Press}},
  \bibinfo{author}{\bibfnamefont{T.~D.} \bibnamefont{Ladd}},
  \bibinfo{author}{\bibfnamefont{B.}~\bibnamefont{Zhang}}, \bibnamefont{and}
  \bibinfo{author}{\bibfnamefont{Y.}~\bibnamefont{Yamamoto}},
  \bibinfo{journal}{Nature} \textbf{\bibinfo{volume}{456}},
  \bibinfo{pages}{218} (\bibinfo{year}{2008}).

\bibitem[{\citenamefont{Kamada et~al.}(2001)\citenamefont{Kamada, Gotoh,
  Temmyo, Takagahara, and Ando}}]{Kamada2001}
\bibinfo{author}{\bibfnamefont{H.}~\bibnamefont{Kamada}},
  \bibinfo{author}{\bibfnamefont{H.}~\bibnamefont{Gotoh}},
  \bibinfo{author}{\bibfnamefont{J.}~\bibnamefont{Temmyo}},
  \bibinfo{author}{\bibfnamefont{T.}~\bibnamefont{Takagahara}},
  \bibnamefont{and} \bibinfo{author}{\bibfnamefont{H.}~\bibnamefont{Ando}},
  \bibinfo{journal}{Phys. Rev. Lett.} \textbf{\bibinfo{volume}{87}},
  \bibinfo{pages}{246401} (\bibinfo{year}{2001}).

\bibitem[{\citenamefont{Zrenner et~al.}(2002)\citenamefont{Zrenner, Beham,
  Stufler, Findeis, Bichler, and Abstreiter}}]{Zrenner2002}
\bibinfo{author}{\bibfnamefont{A.}~\bibnamefont{Zrenner}},
  \bibinfo{author}{\bibfnamefont{E.}~\bibnamefont{Beham}},
  \bibinfo{author}{\bibfnamefont{S.}~\bibnamefont{Stufler}},
  \bibinfo{author}{\bibfnamefont{F.}~\bibnamefont{Findeis}},
  \bibinfo{author}{\bibfnamefont{M.}~\bibnamefont{Bichler}}, \bibnamefont{and}
  \bibinfo{author}{\bibfnamefont{G.}~\bibnamefont{Abstreiter}},
  \bibinfo{journal}{Nature} \textbf{\bibinfo{volume}{418}},
  \bibinfo{pages}{612} (\bibinfo{year}{2002}).

\bibitem[{\citenamefont{Ramsay et~al.}(2010)\citenamefont{Ramsay, Gopal,
  Gauger, Nazir, Lovett, Fox, and Skolnick}}]{Ramsay2010}
\bibinfo{author}{\bibfnamefont{A.~J.} \bibnamefont{Ramsay}},
  \bibinfo{author}{\bibfnamefont{A.~V.} \bibnamefont{Gopal}},
  \bibinfo{author}{\bibfnamefont{E.~M.} \bibnamefont{Gauger}},
  \bibinfo{author}{\bibfnamefont{A.}~\bibnamefont{Nazir}},
  \bibinfo{author}{\bibfnamefont{B.~W.} \bibnamefont{Lovett}},
  \bibinfo{author}{\bibfnamefont{A.~M.} \bibnamefont{Fox}}, \bibnamefont{and}
  \bibinfo{author}{\bibfnamefont{M.~S.} \bibnamefont{Skolnick}},
  \bibinfo{journal}{Phys. Rev. Lett.} \textbf{\bibinfo{volume}{104}},
  \bibinfo{pages}{017402} (\bibinfo{year}{2010}).

\bibitem[{\citenamefont{Gaetan et~al.}(2009)\citenamefont{Gaetan,
  Miroshnychenko, Wilk, Chotia, Viteau, Comparat, Pillet, Browaeys, and
  Grangier}}]{Browaeys2009}
\bibinfo{author}{\bibfnamefont{A.}~\bibnamefont{Gaetan}},
  \bibinfo{author}{\bibfnamefont{Y.}~\bibnamefont{Miroshnychenko}},
  \bibinfo{author}{\bibfnamefont{T.}~\bibnamefont{Wilk}},
  \bibinfo{author}{\bibfnamefont{A.}~\bibnamefont{Chotia}},
  \bibinfo{author}{\bibfnamefont{M.}~\bibnamefont{Viteau}},
  \bibinfo{author}{\bibfnamefont{D.}~\bibnamefont{Comparat}},
  \bibinfo{author}{\bibfnamefont{P.}~\bibnamefont{Pillet}},
  \bibinfo{author}{\bibfnamefont{A.}~\bibnamefont{Browaeys}}, \bibnamefont{and}
  \bibinfo{author}{\bibfnamefont{P.}~\bibnamefont{Grangier}},
  \bibinfo{journal}{Nature Phys.} \textbf{\bibinfo{volume}{5}},
  \bibinfo{pages}{115} (\bibinfo{year}{2009}).

\bibitem[{\citenamefont{Vitanov et~al.}(2001)\citenamefont{Vitanov,
  Fleischhauer, Shore, and Bergmann}}]{Vitanov2001}
\bibinfo{author}{\bibfnamefont{N.~V.} \bibnamefont{Vitanov}},
  \bibinfo{author}{\bibfnamefont{M.}~\bibnamefont{Fleischhauer}},
  \bibinfo{author}{\bibfnamefont{B.~W.} \bibnamefont{Shore}}, \bibnamefont{and}
  \bibinfo{author}{\bibfnamefont{K.}~\bibnamefont{Bergmann}},
  \bibinfo{journal}{Adv. At.,Mol., Opt. Phys.} \textbf{\bibinfo{volume}{46}},
  \bibinfo{pages}{55} (\bibinfo{year}{2001}).

\bibitem[{\citenamefont{Garwood and DelaBarre}(2001)}]{Garwood2001}
\bibinfo{author}{\bibfnamefont{M.}~\bibnamefont{Garwood}} \bibnamefont{and}
  \bibinfo{author}{\bibfnamefont{L.}~\bibnamefont{DelaBarre}},
  \bibinfo{journal}{Journal of Magnetic Resonance}
  \textbf{\bibinfo{volume}{153}}, \bibinfo{pages}{155 } (\bibinfo{year}{2001}).

\bibitem[{\citenamefont{Petta et~al.}(2010)\citenamefont{Petta, Lu, and
  Gossard}}]{Petta2010}
\bibinfo{author}{\bibfnamefont{J.~R.} \bibnamefont{Petta}},
  \bibinfo{author}{\bibfnamefont{H.}~\bibnamefont{Lu}}, \bibnamefont{and}
  \bibinfo{author}{\bibfnamefont{A.~C.} \bibnamefont{Gossard}},
  \bibinfo{journal}{Science} \textbf{\bibinfo{volume}{327}},
  \bibinfo{pages}{669} (\bibinfo{year}{2010}).

\bibitem[{\citenamefont{Melinger et~al.}(1994)\citenamefont{Melinger, Gandhi,
  Hariharan, Goswami, and Warren}}]{Melinger1994}
\bibinfo{author}{\bibfnamefont{J.~S.} \bibnamefont{Melinger}},
  \bibinfo{author}{\bibfnamefont{S.~R.} \bibnamefont{Gandhi}},
  \bibinfo{author}{\bibfnamefont{A.}~\bibnamefont{Hariharan}},
  \bibinfo{author}{\bibfnamefont{D.}~\bibnamefont{Goswami}}, \bibnamefont{and}
  \bibinfo{author}{\bibfnamefont{W.~S.} \bibnamefont{Warren}},
  \bibinfo{journal}{The Journal of Chemical Physics}
  \textbf{\bibinfo{volume}{101}}, \bibinfo{pages}{6439} (\bibinfo{year}{1994}).

\bibitem[{\citenamefont{Maas et~al.}(1999)\citenamefont{Maas, Rella, Antoine,
  Toma, and Noordam}}]{Maas99Rbadiabatic}
\bibinfo{author}{\bibfnamefont{D.~J.} \bibnamefont{Maas}},
  \bibinfo{author}{\bibfnamefont{C.~W.} \bibnamefont{Rella}},
  \bibinfo{author}{\bibfnamefont{P.}~\bibnamefont{Antoine}},
  \bibinfo{author}{\bibfnamefont{E.~S.} \bibnamefont{Toma}}, \bibnamefont{and}
  \bibinfo{author}{\bibfnamefont{L.~D.} \bibnamefont{Noordam}},
  \bibinfo{journal}{Physical Review A} \textbf{\bibinfo{volume}{59}},
  \bibinfo{pages}{1374} (\bibinfo{year}{1999}).

\bibitem[{\citenamefont{Fleischhauer et~al.}(2005)\citenamefont{Fleischhauer,
  Imamoglu, and Marangos}}]{Fleischhauer2005}
\bibinfo{author}{\bibfnamefont{M.}~\bibnamefont{Fleischhauer}},
  \bibinfo{author}{\bibfnamefont{A.}~\bibnamefont{Imamoglu}}, \bibnamefont{and}
  \bibinfo{author}{\bibfnamefont{J.~P.} \bibnamefont{Marangos}},
  \bibinfo{journal}{Rev. Mod. Phys.} \textbf{\bibinfo{volume}{77}},
  \bibinfo{pages}{633} (\bibinfo{year}{2005}).

\bibitem[{\citenamefont{Greilich et~al.}(2006)\citenamefont{Greilich, Yakovlev,
  Shabaev, Efros, Yugova, Oulton, Stavarache, Reuter, Wieck, and
  Bayer}}]{Greilich2006}
\bibinfo{author}{\bibfnamefont{A.}~\bibnamefont{Greilich}},
  \bibinfo{author}{\bibfnamefont{D.~R.} \bibnamefont{Yakovlev}},
  \bibinfo{author}{\bibfnamefont{A.}~\bibnamefont{Shabaev}},
  \bibinfo{author}{\bibfnamefont{A.~L.} \bibnamefont{Efros}},
  \bibinfo{author}{\bibfnamefont{I.~A.} \bibnamefont{Yugova}},
  \bibinfo{author}{\bibfnamefont{R.}~\bibnamefont{Oulton}},
  \bibinfo{author}{\bibfnamefont{V.}~\bibnamefont{Stavarache}},
  \bibinfo{author}{\bibfnamefont{D.}~\bibnamefont{Reuter}},
  \bibinfo{author}{\bibfnamefont{A.}~\bibnamefont{Wieck}}, \bibnamefont{and}
  \bibinfo{author}{\bibfnamefont{M.}~\bibnamefont{Bayer}},
  \bibinfo{journal}{Science} \textbf{\bibinfo{volume}{313}},
  \bibinfo{pages}{341} (\bibinfo{year}{2006}).

\bibitem[{\citenamefont{Eble et~al.}(2009)\citenamefont{Eble, Testelin,
  Desfonds, Bernardot, Balocchi, Amand, Miard, Lema\^\i{}tre, Marie, and
  Chamarro}}]{Eble2009}
\bibinfo{author}{\bibfnamefont{B.}~\bibnamefont{Eble}},
  \bibinfo{author}{\bibfnamefont{C.}~\bibnamefont{Testelin}},
  \bibinfo{author}{\bibfnamefont{P.}~\bibnamefont{Desfonds}},
  \bibinfo{author}{\bibfnamefont{F.}~\bibnamefont{Bernardot}},
  \bibinfo{author}{\bibfnamefont{A.}~\bibnamefont{Balocchi}},
  \bibinfo{author}{\bibfnamefont{T.}~\bibnamefont{Amand}},
  \bibinfo{author}{\bibfnamefont{A.}~\bibnamefont{Miard}},
  \bibinfo{author}{\bibfnamefont{A.}~\bibnamefont{Lema\^\i{}tre}},
  \bibinfo{author}{\bibfnamefont{X.}~\bibnamefont{Marie}}, \bibnamefont{and}
  \bibinfo{author}{\bibfnamefont{M.}~\bibnamefont{Chamarro}},
  \bibinfo{journal}{Phys. Rev. Lett.} \textbf{\bibinfo{volume}{102}},
  \bibinfo{pages}{146601} (\bibinfo{year}{2009}).

\bibitem[{\citenamefont{Eastham and Phillips}(2009)}]{Eastham2009}
\bibinfo{author}{\bibfnamefont{P.~R.} \bibnamefont{Eastham}} \bibnamefont{and}
  \bibinfo{author}{\bibfnamefont{R.~T.} \bibnamefont{Phillips}},
  \bibinfo{journal}{Phys. Rev. B} \textbf{\bibinfo{volume}{79}},
  \bibinfo{pages}{165303} (\bibinfo{year}{2009}).

\bibitem[{\citenamefont{Urbaszek et~al.}(2003)\citenamefont{Urbaszek,
  Warburton, Karrai, Gerardot, Petroff, and Garcia}}]{urbaszek2003}
\bibinfo{author}{\bibfnamefont{B.}~\bibnamefont{Urbaszek}},
  \bibinfo{author}{\bibfnamefont{R.~J.} \bibnamefont{Warburton}},
  \bibinfo{author}{\bibfnamefont{K.}~\bibnamefont{Karrai}},
  \bibinfo{author}{\bibfnamefont{B.~D.} \bibnamefont{Gerardot}},
  \bibinfo{author}{\bibfnamefont{P.~M.} \bibnamefont{Petroff}},
  \bibnamefont{and} \bibinfo{author}{\bibfnamefont{J.~M.}
  \bibnamefont{Garcia}}, \bibinfo{journal}{Phys. Rev. Lett.}
  \textbf{\bibinfo{volume}{90}}, \bibinfo{pages}{247403}
  (\bibinfo{year}{2003}).

\bibitem[{\citenamefont{Paillard et~al.}(2001)}]{Marie01}
\bibinfo{author}{\bibfnamefont{M.}~\bibnamefont{Paillard}}
  \bibnamefont{et~al.}, \bibinfo{journal}{Phys. Rev. Lett.}
  \textbf{\bibinfo{volume}{86}}, \bibinfo{pages}{1634} (\bibinfo{year}{2001}).

\bibitem[{\citenamefont{Ates et~al.}(2009)\citenamefont{Ates, Ulrich, Ulhaq,
  Reitzenstein, Loffler, Hofling, Forchel, and Michler}}]{Michler2009}
\bibinfo{author}{\bibfnamefont{S.}~\bibnamefont{Ates}},
  \bibinfo{author}{\bibfnamefont{S.~M.} \bibnamefont{Ulrich}},
  \bibinfo{author}{\bibfnamefont{A.}~\bibnamefont{Ulhaq}},
  \bibinfo{author}{\bibfnamefont{S.}~\bibnamefont{Reitzenstein}},
  \bibinfo{author}{\bibfnamefont{A.}~\bibnamefont{Loffler}},
  \bibinfo{author}{\bibfnamefont{S.}~\bibnamefont{Hofling}},
  \bibinfo{author}{\bibfnamefont{A.}~\bibnamefont{Forchel}}, \bibnamefont{and}
  \bibinfo{author}{\bibfnamefont{P.}~\bibnamefont{Michler}},
  \bibinfo{journal}{Nat. Photon.} \textbf{\bibinfo{volume}{3}},
  \bibinfo{pages}{724} (\bibinfo{year}{2009}).

\bibitem[{tun()}]{tunnel}
\bibinfo{note}{Should $\tau_\text{in}$ be of the same order as the radiative
  recombination time of the X$^0$ of about 1ns, both X$^0$ and X$^-$ would show
  comparable intensities for non-resonant excitation. As the X$^-$ is under
  these conditions about 15 times more intense than the X$^0$ we estimate
  $\tau_\text{in} \approx 1ns/15 \approx 70ps$}.

\bibitem[{\citenamefont{Dalgarno et~al.}(2008)\citenamefont{Dalgarno, Smith,
  McFarlane, Gerardot, Karrai, Badolato, Petroff, and
  Warburton}}]{Dalgarno2008b}
\bibinfo{author}{\bibfnamefont{P.~A.} \bibnamefont{Dalgarno}},
  \bibinfo{author}{\bibfnamefont{J.~M.} \bibnamefont{Smith}},
  \bibinfo{author}{\bibfnamefont{J.}~\bibnamefont{McFarlane}},
  \bibinfo{author}{\bibfnamefont{B.~D.} \bibnamefont{Gerardot}},
  \bibinfo{author}{\bibfnamefont{K.}~\bibnamefont{Karrai}},
  \bibinfo{author}{\bibfnamefont{A.}~\bibnamefont{Badolato}},
  \bibinfo{author}{\bibfnamefont{P.~M.} \bibnamefont{Petroff}},
  \bibnamefont{and} \bibinfo{author}{\bibfnamefont{R.~J.}
  \bibnamefont{Warburton}}, \bibinfo{journal}{Phys. Rev. B}
  \textbf{\bibinfo{volume}{77}}, \bibinfo{pages}{245311}
  (\bibinfo{year}{2008}).

\bibitem[{\citenamefont{Chatel et~al.}(2003)\citenamefont{Chatel, Degert,
  Stock, and Girard}}]{chatelNa03PRA}
\bibinfo{author}{\bibfnamefont{B.}~\bibnamefont{Chatel}},
  \bibinfo{author}{\bibfnamefont{J.}~\bibnamefont{Degert}},
  \bibinfo{author}{\bibfnamefont{S.}~\bibnamefont{Stock}}, \bibnamefont{and}
  \bibinfo{author}{\bibfnamefont{B.}~\bibnamefont{Girard}},
  \bibinfo{journal}{Phys. Rev. A} \textbf{\bibinfo{volume}{68}},
  \bibinfo{pages}{041402} (\bibinfo{year}{2003}).

\bibitem[{pul()}]{pulsearea}
\bibinfo{note}{The pulse area $\Theta$ does not depend on the chirp, allowing
  for a direct comparison of Rabi oscillations and adiabatic passage.}

\bibitem[{asy()}]{asy}
\bibinfo{note}{This would also help improving the fit of the Rabi oscillations
  in figure 2a by allowing for an assymetry in the oscillations with respect to
  the 50\% line due a small, residual chirp.}

\bibitem[{hyp()}]{hyper}
\bibinfo{note}{Calculations with secant hyperbolic pulses lead to
  quantitatively very similar conclusions.}

\bibitem[{Coh()}]{Cohen}
\bibinfo{note}{Cohen-Tannoudji, C. Dupont-Roc, J. and Grynberg, G.
  \emph{Atom-Photon Interactions}, (Wiley-Interscience, 1998).}

\bibitem[{\citenamefont{Hioe}(1984)}]{Hioe1984}
\bibinfo{author}{\bibfnamefont{F.~T.} \bibnamefont{Hioe}},
  \bibinfo{journal}{Phys. Rev. A} \textbf{\bibinfo{volume}{30}},
  \bibinfo{pages}{2100} (\bibinfo{year}{1984}).

\bibitem[{\citenamefont{Bennett et~al.}(2009)\citenamefont{Bennett, Patel,
  Nicoll, Ritchie, and Shields}}]{Bennett2009}
\bibinfo{author}{\bibfnamefont{A.~J.} \bibnamefont{Bennett}},
  \bibinfo{author}{\bibfnamefont{R.~B.} \bibnamefont{Patel}},
  \bibinfo{author}{\bibfnamefont{C.}~\bibnamefont{Nicoll}},
  \bibinfo{author}{\bibfnamefont{D.~A.} \bibnamefont{Ritchie}},
  \bibnamefont{and} \bibinfo{author}{\bibfnamefont{A.~J.}
  \bibnamefont{Shields}}, \bibinfo{journal}{Nature Phys.}
  \textbf{\bibinfo{volume}{5}}, \bibinfo{pages}{715} (\bibinfo{year}{2009}).

\bibitem[{\citenamefont{Akopian et~al.}(2006)\citenamefont{Akopian, Lindner,
  Poem, Berlatzky, Avron, Gershoni, Gerardot, and Petroff}}]{Akopian2006}
\bibinfo{author}{\bibfnamefont{N.}~\bibnamefont{Akopian}},
  \bibinfo{author}{\bibfnamefont{N.~H.} \bibnamefont{Lindner}},
  \bibinfo{author}{\bibfnamefont{E.}~\bibnamefont{Poem}},
  \bibinfo{author}{\bibfnamefont{Y.}~\bibnamefont{Berlatzky}},
  \bibinfo{author}{\bibfnamefont{J.}~\bibnamefont{Avron}},
  \bibinfo{author}{\bibfnamefont{D.}~\bibnamefont{Gershoni}},
  \bibinfo{author}{\bibfnamefont{B.~D.} \bibnamefont{Gerardot}},
  \bibnamefont{and} \bibinfo{author}{\bibfnamefont{P.~M.}
  \bibnamefont{Petroff}}, \bibinfo{journal}{Phys. Rev. Lett.}
  \textbf{\bibinfo{volume}{96}}, \bibinfo{pages}{130501}
  (\bibinfo{year}{2006}).

\bibitem[{\citenamefont{Dousse et~al.}(2010)\citenamefont{Dousse, Suffczynski,
  Beveratos, Krebs, Lemaitre, Sagnes, Bloch, Voisin, and
  Senellart}}]{Dousse2010}
\bibinfo{author}{\bibfnamefont{A.}~\bibnamefont{Dousse}},
  \bibinfo{author}{\bibfnamefont{J.}~\bibnamefont{Suffczynski}},
  \bibinfo{author}{\bibfnamefont{A.}~\bibnamefont{Beveratos}},
  \bibinfo{author}{\bibfnamefont{O.}~\bibnamefont{Krebs}},
  \bibinfo{author}{\bibfnamefont{A.}~\bibnamefont{Lemaitre}},
  \bibinfo{author}{\bibfnamefont{I.}~\bibnamefont{Sagnes}},
  \bibinfo{author}{\bibfnamefont{J.}~\bibnamefont{Bloch}},
  \bibinfo{author}{\bibfnamefont{P.}~\bibnamefont{Voisin}}, \bibnamefont{and}
  \bibinfo{author}{\bibfnamefont{P.}~\bibnamefont{Senellart}},
  \bibinfo{journal}{Nature} \textbf{\bibinfo{volume}{466}},
  \bibinfo{pages}{217} (\bibinfo{year}{2010}).

\bibitem[{\citenamefont{Calarco et~al.}(2003)\citenamefont{Calarco, Datta,
  Fedichev, Pazy, and Zoller}}]{Zoller2003}
\bibinfo{author}{\bibfnamefont{T.}~\bibnamefont{Calarco}},
  \bibinfo{author}{\bibfnamefont{A.}~\bibnamefont{Datta}},
  \bibinfo{author}{\bibfnamefont{P.}~\bibnamefont{Fedichev}},
  \bibinfo{author}{\bibfnamefont{E.}~\bibnamefont{Pazy}}, \bibnamefont{and}
  \bibinfo{author}{\bibfnamefont{P.}~\bibnamefont{Zoller}},
  \bibinfo{journal}{Phys. Rev. A} \textbf{\bibinfo{volume}{68}},
  \bibinfo{pages}{012310} (\bibinfo{year}{2003}).

\bibitem[{\citenamefont{Troiani et~al.}(2003)\citenamefont{Troiani, Molinari,
  and Hohenester}}]{Hohenester2003}
\bibinfo{author}{\bibfnamefont{F.}~\bibnamefont{Troiani}},
  \bibinfo{author}{\bibfnamefont{E.}~\bibnamefont{Molinari}}, \bibnamefont{and}
  \bibinfo{author}{\bibfnamefont{U.}~\bibnamefont{Hohenester}},
  \bibinfo{journal}{Phys. Rev. Lett.} \textbf{\bibinfo{volume}{90}},
  \bibinfo{pages}{206802} (\bibinfo{year}{2003}).

\bibitem[{\citenamefont{Chen et~al.}(2004)\citenamefont{Chen, Piermarocchi,
  Sham, Gammon, and Steel}}]{Steel2004}
\bibinfo{author}{\bibfnamefont{P.}~\bibnamefont{Chen}},
  \bibinfo{author}{\bibfnamefont{C.}~\bibnamefont{Piermarocchi}},
  \bibinfo{author}{\bibfnamefont{L.~J.} \bibnamefont{Sham}},
  \bibinfo{author}{\bibfnamefont{D.}~\bibnamefont{Gammon}}, \bibnamefont{and}
  \bibinfo{author}{\bibfnamefont{D.~G.} \bibnamefont{Steel}},
  \bibinfo{journal}{Phys. Rev. B} \textbf{\bibinfo{volume}{69}},
  \bibinfo{pages}{075320} (\bibinfo{year}{2004}).

\end{thebibliography}

\end{document}